\documentclass{svproc}
\usepackage{url}
\usepackage{amsmath,amssymb}
\usepackage{bm}
\usepackage{graphicx}
\usepackage{float}

\begin{document}
\mainmatter              % start of a contribution
\title{Measurement Optimization under Uncertainty using Deep Reinforcement Learning}
\titlerunning{Measurement Optimization}  % abbreviated title (for running head)
%                                     also used for the TOC unless
%                                     \toctitle is used
%
\author{Amin Jabini\inst{1} \and Erik A. Johnson\inst{1}}
\authorrunning{Amin Jabini and Erik A. Johnson} % abbreviated author list (for running head)
%
%%%% list of authors for the TOC (use if author list has to be modified)
\tocauthor{Amin Jabini and Erik A. Johnson}
\institute{$^1$ University of Southern California, Los Angeles CA 90089, USA,\\
emails: \email{jabini@usc.edu and JohnsonE@usc.edu}
}

\maketitle              % typeset the title of the contribution

\begin{abstract}
Optimal sensor placement enhances the efficiency of a variety of applications for monitoring dynamical systems. It has been established that deterministic solutions to the sensor placement problem are insufficient due to the many uncertainties in system input and parameters that affect system response sensor measurements. Accounting for the uncertainties in this typically expensive optimization is challenging due to computational intractability. This study proposes a stochastic environment in the form of a Markov Decision Process and a sensor placement agent that aims to maximize the information gain from placing a particular number of sensors of different types within the system. The agent is trained to maximize its reward based on an information-theoretic reward function. To verify the efficacy,  the approach is applied to place a set of heterogeneous sensors in a shear building model. This methodology can be used to accommodate uncertainties in the sensor placement problem in real-world systems.
% We would like to encourage you to list your keywords within
% the abstract section using the \keywords{...} command.
\end{abstract}
\section{Introduction}
Infrastructure management is based on the information gathered from the system through measurements \cite{MADANAT199377}. Traditionally, this information has been gathered through visual inspection. The frequency and quality of inspections form a decision-making problem \cite{thoft1987optimal,yang1975inspection}. As sensors have become more affordable, data acquisition systems have been replacing the inspection process. In most cases, measurements are collected using sensor networks composed of different types of sensors. The design problem now consists of determining the number and location of each type of sensor. The total number of sensors is usually constrained by budget, but the sensor types and locations must be chosen in a way that maximizes effectiveness. For example, in structural health monitoring applications, the sensor configuration should adequately capture the structure vibration. The sensor placement decisions form an optimization to find the most informative sensor configuration \cite{krause2008optimizing}. Similar optimizations are also involved in designing monitoring systems for other domains, such as water distribution systems \cite{krause2008efficient} and flame detection systems \cite{zhen2019mathematical}, as well as wearable sensors \cite{atallah2010sensor}. 

Several evaluation functions have been used to measure the effectiveness of a sensor configuration. The quality of estimation based on each sensor layout has been used by some researchers as a measure; the norm of the resulting covariance matrix of the estimated parameters \cite{udwadia1994methodology}, the expected information gain \cite{huan2013simulation}, various norms of the Fisher information matrix \cite{atallah2010sensor}, and the observability Gramian matrix \cite{udwadia1994methodology} are in this category. To optimize the objective function, various frameworks have been applied, such as forward and backward sequential optimization \cite{papadimitriou2004optimal}, swarm intelligence \cite{yi2015health}, and Nelder-Mead Nonlinear Simplex optimization (MMNS) \cite{huan2013simulation}. A common drawback of these methods involves scaling issues: because sensor placement is combinatorial in nature, as the number of candidate locations increases in a complex system, these methods become ineffective. In addition, most objective functions do not consider the inherent uncertainties in the problem. Udwadia \cite{udwadia1994methodology} showed that the optimal sensor configuration depends on the nature of the system, the model parameter values, and the nature and location of the input forces. Considering these uncertainties makes the current approaches computationally intractable. This study instead poses a stochastic solution based on deep reinforcement learning.

Reinforcement Learning (RL) is the formulation of an agent in an embedded environment that seeks an optimal policy to maximize its reward \cite{sutton2018reinforcement}. The policy function is a mapping between the state of the environment and the action space. The agent learns the best mapping by gathering experience from the environment and improves the policy using the reward it collects for actions. While any function can be used as an approximation for the policy function, deep learning models, such as neural networks with many layers, have proven to be powerful in expressing complex mappings. The use of deep learning models in RL, which is called deep reinforcement learning (DRL), resulted in groundbreaking achievements in robotics \cite{levine2016end}, control \cite{levine2018reinforcement}, and computer games \cite{mnih2013playing,mnih2015huma}, and has revolutionized many decision-making problems and optimizations in resource optimization \cite{mao2016resource}, chip design \cite{mirhoseini2020chip},
% traffic control \cite{arel2010reinforcement}
and system maintenance planning \cite{ANDRIOTIS2019106483}.

The DRL framework has the capability to account for the mentioned uncertainties in the sensor placement problem. Furthermore, the formulation can be adapted without limitation to different settings, such as different sensor types. Moreover, the function approximation in the policy enables scalability of the solution to systems with higher dimensions of state and action spaces. Methods for adaptive search of the design space can be utilized to make the training procedure more efficient. For example, simulated annealing can help the agent explore the sensor configuration space broadly at first and gradually exploit the experience gathered to move toward more rewarding configurations. 

This study proposes a DRL-based approach for heterogeneous sensor placement under parameter and input uncertainty. The reward function is based on information gain, which quantifies the change in entropy of parameter uncertainties. In this study, the number of sensors to be used is selected as a constraint, as is common in sensor network design. To maximize information gain, the problem is to select sensors from different types and assign them to candidate locations across the structural system. 

\section{Problem Formulation}
Consider the placement of $m$ sensors, to be selected from $p$ sensor types, among $n$ candidate locations. The solution considers the uncertainty in the input excitation and the parameter values. In this section, a brief summary of reinforcement learning is reviewed, the MDP for the sensor placement is formulated, and the reward function is discussed.

\subsection{Reinforcement learning}
Reinforcement learning is the problem of learning some optimal behavior in an environment based on interacting with it over multiple iterations. The problem is defined as a Markov Decision Process (MDP). Herein, a discrete MDP defined by the tuple $(\mathcal{S}, \mathcal{A}, \mathcal{P}, r, \gamma, \bm{s}_0)$ is used to model the sensor placement problem where $\mathcal{S}$ is a set of discrete states, $\mathcal{A}$ is a set of discrete actions, $\mathcal{P}$ is the state transition probability, $r$ is a $\mathcal{S}\times \mathcal{A} \rightarrow \mathbb{R}$ reward function, $\gamma$ is the future discount factor, and $\bm{s}_0 \in \mathcal{S}$ is the initial state. The goal of the agent is to maximize its expected future reward:
\begin{equation}
  J(\bm{\phi}) = \mathop{\mathbb{E}}\left[\sum_{t=1}^{T} \gamma^{t-1} r(\bm{s}_t, \bm{a}_t) \right]
  \label{eq: total_reward}
\end{equation}
where $\bm{\phi}$ parameterizes the agent's policy function, $T$ is the horizon, and $\bm{s}_t$ and $\bm{a}_t$ are is the environment state and the agent's action, respectively, at time $t$. Various classes of algorithms have been proposed for RL. Herein, the double Q-network \cite{NIPS2010_091d584f}, an action-value algorithm for finding a near optimal policy, is used as discussed in the next section. 

\subsection{Double Q-network}
\begin{figure}[htp]
\centering
    \vspace{-2em}%
    \includegraphics[width=8.5cm]{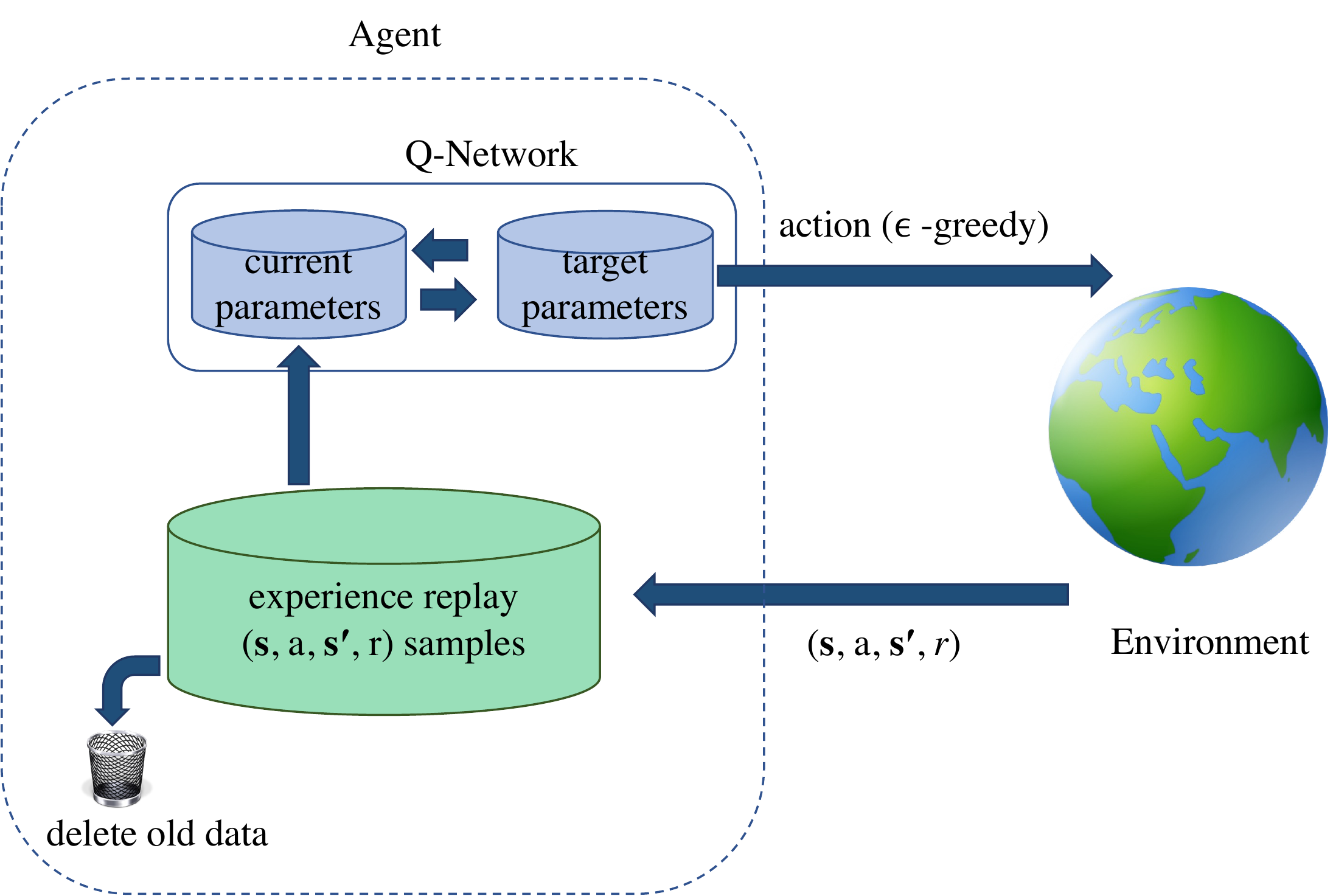}
    \caption{A schematic of the Double Q-learning with experience replay}
    \label{fig:DDQN}
    \vspace{-2em}%
\end{figure}

In Q-learning methods \cite{sutton2018reinforcement}, the goal is to estimate the average reward from each state-action pair instead of learning the policy function directly. This average future reward is denoted by:
\begin{equation}
  Q(\bm{s}_i, \bm{a}_i) = \mathop{\mathbb{E}}\left[\sum_{t=t_i}^{T} \gamma^{t-1} r(\bm{s}_t, \bm{a}_t) \bigg\vert \bm{s}_{t_i}=\bm{s}_i, \bm{a}_{t_i}=\bm{a}_i \right]
  \label{eq: Q-value}
\end{equation}
The Q-values can be represented by a table in the simplest form, or as a parameterized function $Q_\phi(\bm{s},\bm{a})$ that is captured by a neural network model. The best policy, $\pi(\bm{s}_i)$ at state $\bm{s}_i$, is simply the action that has the highest Q-value for that state:
\begin{equation}
  \pi(\bm{s}_i) = \arg\max_{\bm{a}} Q_\phi(\bm{s}_i, \bm{a})
  \label{eq: argmax_policy}
\end{equation}
The Q-learning algorithm in its simplest form may suffer from convergence problems. To improve its convergence, several strategies have been proposed in the literature. One issue is that when the training starts and the Q-function is initialized with random weights, the Q-function is not yet sufficiently reliable to choose the best actions because the agent needs to more widely explore the environment. Therefore, scheduling schemes like $\epsilon$-greedy \cite{sutton2018reinforcement} are utilized to initially prefer exploration over exploitation. Another issue is that the dataset to which the Q-function $Q_\phi$ is being fit is itself evolving as data is collected from the environment.  Because the agent's action is selected based on that Q-function, this means that the agent's action also evolves as data is collected.  As a result, if the Q-function is updated in each step of the training, convergence is not guaranteed.  However, it has been shown \cite{NIPS2010_091d584f} that using two Q-functions --- $Q_\phi$, which is called the target function, that is updated only periodically and $Q_{\phi^\prime}$, which is called the local function, that is updated more frequently (and, thus, the former is a earlier version of the latter) --- mitigates this convergence issue.
Further, to make the training more efficient, an experience replay, which is a memory buffer that consists of previously collected experience tuples in the form of $(\bm{s}_i, \bm{a}_i, \bm{s}_{i+1}, r_i)$, is used \cite{Adam2012Exper}. This buffer allows the separation of data collection and training and regularly replaces the old data with the new experience. The schematic of the double Q-learning with experience replay is shown in Fig \ref{fig:DDQN}.
\subsection{Sensor placement MDP}
The MDP for the heterogenous sensor placement problem is as follows: $\mathcal{S}$ is the set of sensor configurations, each element of which is represented by a vector of size $n p$ consisting of zeros and ones to represent the sensor type and their locations. $\mathcal{A}$ is the set of all candidate sensor types and locations; each element is an integer in $[1, n p]$.
At each step, the agent selects a new location and the associated sensor type for the next sensor --- together called an action $a$ (now a scalar) --- and the environment transforms to a new state. The agent receives a reward $r$ based on the selected action. Each episode contains $m$ time steps, starting from the initial state $s_0$ being a zero vector, i.e., no sensors, and then adding one sensor in each ``time step.''
\subsection{Information gain as reward function}
This study uses an information-theoretic reward function that is quantified by the change in information entropy of model parameters with the new selected measurement channel \cite{pant2018information}. For a structural system with parameters $\bm{\theta}$ the state equation is defined by the 
\begin{eqnarray}
   \dot{\mathbf{x}}(t) = \mathbf{f}\left(\mathbf{x}(t), \bm{\theta}, \mathbf{u}(t)\right)
  \label{eq: state_eq}
\end{eqnarray}
in which $\mathbf{x}$ is the state vector, and $\mathbf{u}$ is the system input. The observation equation is
\begin{eqnarray}
   \mathbf{y}(t) = \mathbf{C} \mathbf{x}(t) + \mathbf{v}(t)
  \label{eq: observation_eq}
\end{eqnarray}
where \(\mathbf{y}\) is the measured response, $\mathbf{C}$ is the observation matrix, and $\mathbf{v}$ is a zero-mean Gaussian measurement noise with covariance \(\mathbf{R}\). The time-discretized integral of the Fisher information matrix \(\mathbf{F}\) is \cite{udwadia1994methodology,ebrahimian2019information}:
\begin{eqnarray}
   \mathbf{F} \approx \sum_{i=0}^{k} \left[ \frac{\partial \mathbf{y}(i\Delta t)}{\partial \bm{\theta}}\right] ^\mathrm{T}_{\bm{\theta}=\hat{\bm{\theta}}}  \mathbf{R}^{-1} \left[ \frac{\partial \mathbf{y}(i\Delta t)}{\partial \bm{\theta}}\right] _{\bm{\theta}=\hat{\bm{\theta}}}
  \label{eq: fisher}
\end{eqnarray}
where $\hat{\bm{\theta}}$ is the \emph{maximum a posteriori} (MAP) estimate of $\theta$. Ebrahimian et al. \cite{ebrahimian2019information} proved that when the parameters' prior estimate follows a normal distribution and the model uncertainty is negligible, the parameter information gain can be computed as 
\begin{eqnarray}
  \Delta H(\bm{\theta}) =  \frac{1}{2} \log(|\mathbf{F}+\mathbf{P}_{0}^{-1}|)-\frac{1}{2}\log \left(|\mathbf{P}_{0}^{-1}|\right)
  \label{eq: delta_h}
\end{eqnarray}
where $\Delta H(\cdot)$ is the gain in information entropy, $|\cdot|$ is the matrix  determinant, and $\mathbf{P}_{0}$ is the initial covariance matrix of parameters. Equation (\ref{eq: delta_h}) shows that the information gain depends on the probability distribution of the parameters as well as the input that induces the response  \(\mathbf{y}\). Herein, probability distributions are assigned instead of using a single deterministic value. The input is modeled accordingly using a stochastic model. In each episode, the information gain is calculated using samples of parameter values and the input time history, both of which are generated using probabilistic models. The resulting $np \times n_\theta$ information gain matrix is then normalized by its maximum element; the reward function for a particular action corresponds to the one-norm (sum) of the corresponding row of this matrix.
\section{Validation}
The proposed framework is applied to a synthetic example of a four-degree-of-freedom shear building model. The decision is to choose three sensors from accelerometers, velocity sensors and drift measurements with different covariance matrices. The diagonal elements of the covariance matrix $\mathbf{R}$ are $10^{-3}$\,m\textsuperscript{2}/s\textsuperscript{4}, $10^{-5}$\,m\textsuperscript{2}/s\textsuperscript{2} and $10^{-6}$\,m\textsuperscript{2} for the acceleration, interstory drift velocity, and interstory drift sensors, respectively. The probability distributions for the model parameters are considered to be Gaussian: the mean values of the story stiffnesses are 175, 175, 140 and 140 $\mbox{MN/m}$, respectively (starting with the first story), and the damping coefficients are
1.75, 1.75, 1.4, and 1.4 \mbox{MN$\cdot$s/m}, respectively. The coefficient of variation is 0.2 is for all parameters. 
The fundamental mode of the structure using the mean parameter values has a 0.45 second period and a modal damping ratio of 7\%. 
% The matrix $\mathbf{C}$ in Eq. \ref{eq: observation_eq} for measuring the absolute acceleration, inter-story drift and velocity is:
% \begin{eqnarray}
%   \mathbf{C}= \left[\begin{array}{cc}
%     \mathbf{D} & 0 \\
%     0 & \mathbf{D} \\
%     -\mathbf{M}^{-1} \mathbf{K} & -\mathbf{M}^{-1} \mathbf{C}
%     \end{array}\right]
%   \label{eq: C_matrix}
% \end{eqnarray}
% where 
% \begin{eqnarray}
%   \mathbf{D}=\left[\begin{array}{cccc}
%         1 & 0 & 0 & 0 \\
%         -1 & 1 & 0 & 0 \\
%         0 & -1 & 1 & 0 \\
%         0 & 0 & -1 & 1
%         \end{array}\right]
%   \label{eq: D_matrix}
% \end{eqnarray}
\subsection{Ground motion generator}
To model the uncertainties in the input excitation, the Kanai-Tajimi shaping filter \cite{ramallo2002smart} is used with parameters $\omega_g=17 \frac{\text{rad}}{sec}$ and $\zeta_g=0.3$:
\begin{eqnarray}
  F(s) = \frac{2\zeta_g\omega_g s + \omega_g^2}{s^2+2\zeta_g\omega_g s + \omega_g^2}
  \label{eq: kanai_tajimi}
\end{eqnarray}
A Gaussian white noise is passed through this filter to generate random input excitations of 10 second duration and sampling time 0.01 second and then scaled to the a peak ground acceleration of 1.5 m/s\textsuperscript{2}. 
\begin{figure}[b!]
\centering
    \vspace{-2em}%
    \includegraphics[width=8.5cm]{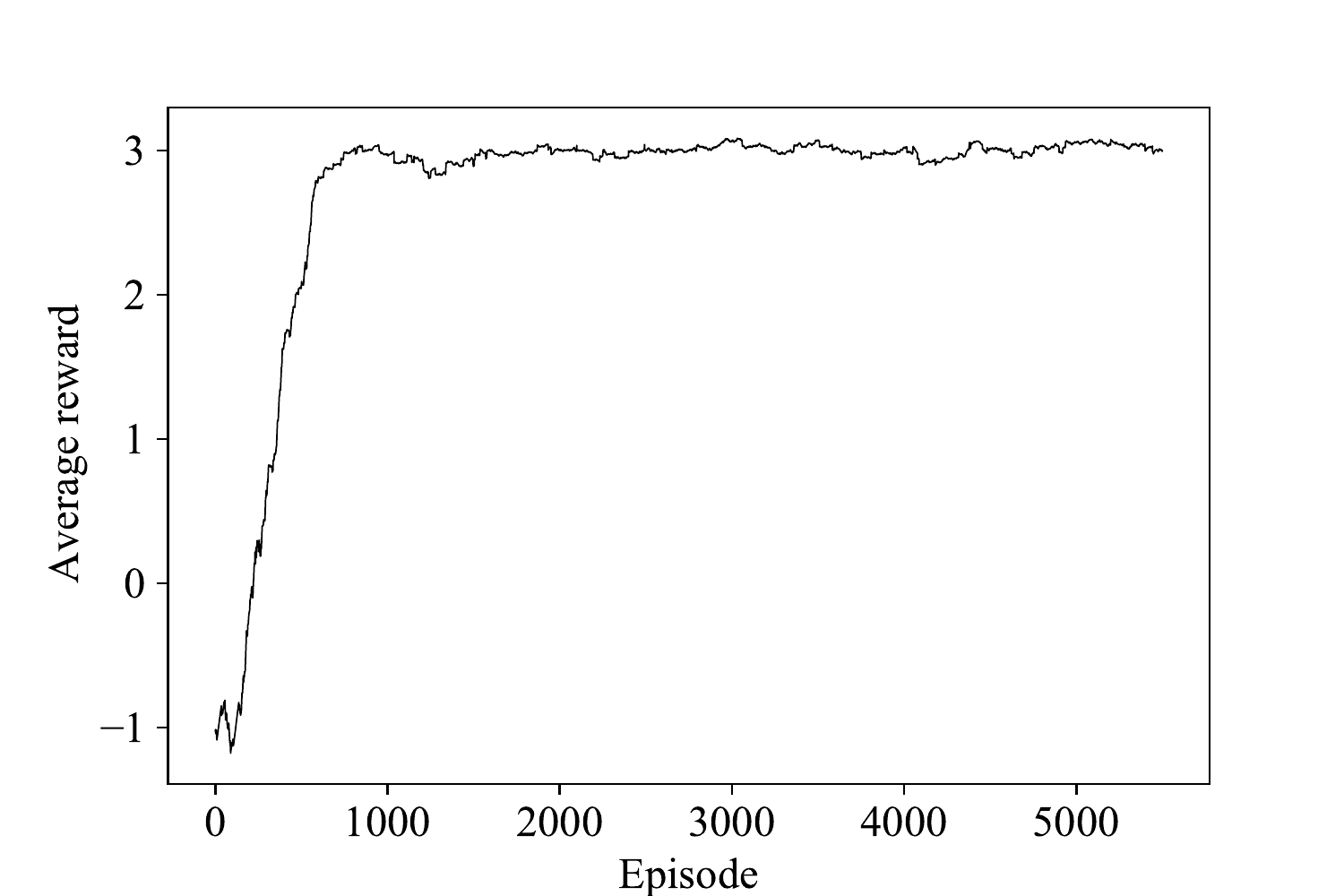}
    \caption{Average episode reward for the agent}
    \label{fig:average_reward}
    \vspace{-2em}%
\end{figure}
\subsection{Results}
The Q-network used in this simulation has three layers, with 12, 6 and 12 node, respectively. The activation function for the first layer is a Rectified Linear Unit (ReLU) and the second layer has a linear activation. The capacity of the experience replay is set to 2000 and the agent is trained for 5500 episodes. The average episode reward is illustrated in Fig \ref{fig:average_reward}. The final policy selects the inter-story drifts of the first and second floors and the drift velocity of the third floor as the best measurement channels.
\section{Conclusion}
A deep reinforcement learning based framework is proposed for sensor placement that has the capability to account for various sensor types, each with different accuracy levels, under input and parameter uncertainty. The function approximation in DRL decreases the sample complexity and enables a more efficient the search over the design space.
\section{Acknowledgement}
The authors acknowledge the financial support of this work by National Science Foundation (NSF) through grant 16-63667, and the first author's Annenberg Fellowship support from the University of Southern California (USC). The authors also acknowledge the Center for Advanced Research Computing (CARC) at the University of Southern California for providing computing resources that have contributed to the research results reported within this publication.  The opinions, findings, and conclusions, or recommendations expressed are those of the author and do not necessarily reflect the views of the NSF or USC.

%
% ---- Bibliography ----
%
\bibliography{biblio} \bibliographystyle{ieeetr}

% \begin{thebibliography}{6}
% %

% \bibitem {smit:wat}
% Smith, T.F., Waterman, M.S.: Identification of common molecular subsequences.
% J. Mol. Biol. 147, 195?197 (1981). \url{doi:10.1016/0022-2836(81)90087-5}

% \bibitem {may:ehr:stein}
% May, P., Ehrlich, H.-C., Steinke, T.: ZIB structure prediction pipeline:
% composing a complex biological workflow through web services.
% In: Nagel, W.E., Walter, W.V., Lehner, W. (eds.) Euro-Par 2006.
% LNCS, vol. 4128, pp. 1148?1158. Springer, Heidelberg (2006).
% \url{doi:10.1007/11823285_121}

% \bibitem {fost:kes}
% Foster, I., Kesselman, C.: The Grid: Blueprint for a New Computing Infrastructure.
% Morgan Kaufmann, San Francisco (1999)

% \bibitem {czaj:fitz}
% Czajkowski, K., Fitzgerald, S., Foster, I., Kesselman, C.: Grid information services
% for distributed resource sharing. In: 10th IEEE International Symposium
% on High Performance Distributed Computing, pp. 181?184. IEEE Press, New York (2001).
% \url{doi: 10.1109/HPDC.2001.945188}

% \bibitem {fo:kes:nic:tue}
% Foster, I., Kesselman, C., Nick, J., Tuecke, S.: The physiology of the grid: an open grid services architecture for distributed systems integration. Technical report, Global Grid
% Forum (2002)

% \bibitem {onlyurl}
% National Center for Biotechnology Information. \url{http://www.ncbi.nlm.nih.gov}

% \end{thebibliography}

\end{document}